\documentstyle[art12]{article}
\def\t2{\tilde t}
\def\u2{\tilde u}
\def\s2{\tilde s}
\begin{document}
\begin{center}
{\large\bf Intrinsic Charm in Proton and $J/\psi$
  Photoproduction\\
    at High Energies}

{\bf V.A. Saleev\\
 Samara State University, 443011 Samara, Russia}
\end{center}

\begin{abstract}
Based on a perturbative theory of quantum chromodynamics and non-relativistic
quark model, associated $J/\psi$ plus open charm photoproduction on charm
quarks in a proton via partonic subprocess $\gamma c\to J/\psi c$ is discussed.
It is shown that the value and energy dependence of the cross section for
such process remarkably depends on the choice of charm distribution
function in a proton. In the region of small $z=E_J/E_{\gamma}<0.2$ the
contribution of the $\gamma c\to J/\psi c$ subprocess in the inelastic
$J/\psi$ photoproduction spectra is larger than the contribution  of the
 photon-gluon fusion subprocess. At the energy range of HERA collider
intrinsic charm contribution in the total inclusive $J/\psi$ photoproduction
cross section may be equal to 4\% of the dominant contribution of
photon-gluon fusion mechanism.
 \end{abstract}
  \begin{section}{Introduction}
  The study of the charm quark distribution function in a proton taiks a
  remarkable interest from the point of view of a investigation of the
  non-perturbative proton wave function \cite{1} as well as from the
  point of view of a calculation of the intrinsic charm quark contribution to
  the processes of charm particle and quarkonium production in
  photon-hadron, hadron-hadron and hadron-nucleus interactions at high
  energies \cite{2,3}.
  The existing data on charm production processes gives us the idea \cite{1,4}
  of a small, but finite (0.3\%--0.5\%), non-perturbative $c\bar c$-component
  in proton wave function. It helps to explain some effects in charm production
  processes which are difficult for understanding in the assumption that
  charm quarks were produced only in hard partonic subprocesses \cite{1}.

 The $J/\psi$ photoproduction process is the best source of information
 about gluon structure function in a proton \cite{5}, especially in the
 region of very small $x$\cite{5b}. In the deep inelastic domain
 $J/\psi$ photoproduction process may be described using so-called colour
  singlet model \cite{6,7}, in which $J/\psi$ is produced in $\gamma g
  \to J/\psi g$ partonic subprocess. Obviously, in the proton
   fragmentation region the contribution of the intrinsic charm in proton
   via $\gamma c\to J/\psi c$ subprocess is too comparably large \cite{8}.
 The partonic subprocess $\gamma c\to J/\psi c$ gives the main contribution
 to associated $J/\psi$ plus open charm photoproduction and may be very
 clean test for charm quark sea in a proton.

 In Sec.2 we shall obtain amplitude, differential and total cross
 section for partonic subprocess $\gamma c\to J/\psi c$ using perturbative
 theory of QCD and non-relativistic quark model. The connection of the
 partonic and measurable cross sections for processes $\gamma p\to J/\psi c X$
 and $\gamma p\to J/\psi X$ is discussed in Sec.3. In that part we also
 present charm quark distribution functions which are used in calculations.
 In Sec.4 we shall calculate $z$-spectra and total cross sections for
 associated $J/\psi$ plus open charm photoproduction process as well as
 the $\gamma c\to J/\psi c$ subprocess contribution to the inclusive
 $J/\psi$ photoproduction cross section at high energies.

 \section{Partonic subprocess $\gamma c\to J/\psi c$}
  In the lowest order of QCD perturbative theory associated $J/\psi$
  plus open charm photoproduction in $\gamma p$-interactions
corresponds to partonic subprocess $\gamma c\to J/\psi c$. It is
described by Feynman diagrams which are shown in Fig.1. The quarkonium
is represented as non-relativistic  quark-untiquark  bound  system  in
singlet colour state with specified mass $M=2m$ (m is c-quark mass) and
spin-parity $J^p=1^-$.

The amplitude of subprocess $\gamma c\to J/\psi c$ can be expressed in
the form:
\begin{equation}
 M=M_1+M_2+M_3+M_4,
\end{equation}
where
\begin{equation}
 M_1=e_qeg^2\bar U(q')\hat\varepsilon_{\gamma}\frac{\hat q'-\hat k+m}
 {(k-q')^2-m^2}T^a\gamma_{\mu}\frac{\delta^{ab}g_{\mu\nu}}{(p-q)^2}
   T^b\hat P\gamma_{\nu}U(q),
\end{equation}
\begin{equation}
 M_2=e_qeg^2\bar U(q')\gamma_{\mu}T^b\hat
P\frac{\delta^{ab}g_{\mu\nu}}{(p+q')^2}
\gamma_{\nu}T^a\frac{\hat k+\hat q+m}
 {(k+q)^2-m^2}
 \hat\varepsilon_{\gamma}U(q),
\end{equation}
\begin{equation}
 M_3=e_qeg^2\bar U(q')\gamma_{\mu}T^a\hat
P\frac{\delta^{ab}g_{\mu\nu}}{(p+q')^2}
 \hat\varepsilon_{\gamma}\frac{\hat p-\hat k+m}
 {(p-k)^2-m^2}
\gamma_{\nu}T^bU(q),
\end{equation}
\begin{equation}
M_4=e_qeg^2\bar U(q')\gamma_{\mu}T^a
\frac{\hat k-\hat p+m} {(p-k)^2-m^2} \hat\varepsilon_{\gamma}\hat P
\frac{\delta^{ab}g_{\mu\nu}}{(p-q)^2}\gamma_{\nu}U(q).
\end{equation}
In these formula:
$$\hat P=\frac{F_c}{\sqrt 2}A\hat\varepsilon_J(\hat p+m),$$
$A=\Psi(0)/\sqrt m$,  $F_c=\delta^{kr}/\sqrt{3}$,  k and  r  are  colour
indexes of charm quarks,
$T^a=\lambda^a/2$, $e_q$ is electrical charge of
$Ó$-quark in units of $e$.  It is well known that $\Psi(0)$, which is
equal to $J/\psi$ wave function at zero point,  can be extracted  in
the lowest  order  of perturbative QCD from the leptonic decay
width of the $J/\psi$:
\begin{equation}
  \Gamma_{ee}=4\pi e_q^2\alpha^2\frac{|\Psi(0)|^2}{m^2}.
\end{equation}
We shall put in our calculation $\Gamma_{ee}=5.4$ KeV \cite{9}.

If we average and sum over spins and  colours  of  initial  and  final
particles, we obtain the expression for square of matrix element:
  \begin{equation}
  |\bar M|^2=\frac{B_{\gamma c}}{m^2}\sum_{j\geq i=1}^{4}K_{ij}
(\tilde s,\tilde t,\tilde u),
   \end{equation}
    where
  $$B_{\gamma c}=\frac{32 \pi^2\alpha_s^2\Gamma_{ee}m}
    {9\alpha},$$
$$\tilde s=\hat s/m^2,\qquad \tilde t=\hat t/m^2,\qquad \tilde  u=\hat
u/m^2,$$
$\hat s, \hat t, \hat u$ are usual Mandelstam variables
and $\hat s+\hat t+\hat u=6m^2$. The explicit analytical formula for
functions $K_{ij}$ have the following forms:


\begin{eqnarray}
      K_{11}&=&-(2\s2\t2-2\s2+\t2^2\u2-4\t2^2-8\t2\u2+14
      \t2+7\u2-106)\nonumber\\
     &&/(4*(\t2^4-4\t2^3+6\t2^2-4\t2+1))
\end{eqnarray}
\begin{eqnarray}
      K_{12}&=&(\s2^3-\s2^2\t2-6\s2^2-\s2\t2^2-2\s2\t2\u2+16
      \s2\t2-\s2\u2^2+8\s2\u2\nonumber\\
      && -28\s2+\t2^3-6\t2^2-\t2\u2
      ^2+8\t2\u2-28\t2+4\u2^2-126\u2+276)/\nonumber\\
      && (4(\s2^2\t2^2-2\s2^2\t2+\s2^2-2\s2\t2^2+4\s2\t2-
      2\s2+\t2^2-2\t2+1))
\end{eqnarray}
\begin{eqnarray}
      K_{13}&=&(\s2^2\t2+\s2^2+2\s2\t2^2+\s2\t2\u2-22\s2\t2+7
      \s2\u2+16\s2+\nonumber\\
&&  2\t2^2\u2-24\t2^2-11\t2\u2+217
 \t2-2\u2^2+41\u2-511)/(2*(\s2\t2^2\u2-4\s2\nonumber\\
&& \t2^2-2\s2\t2\u2+8\s2\t2+\s2\u2-4\s2-\t2^2\u2+4
      \t2^2+2\t2\u2-8\t2-\u2+4))
\end{eqnarray}
\begin{eqnarray}
      K_{14}&=&-(\s2^3+\s2^2\t2-18\s2^2+\s2\t2^2+\s2\t2\u2-16
      \s2\t2-\s2\u2^2-3\s2\u2+\nonumber\\
       &&166\s2-11\t2^2-5\t2\u2+
 115\t2+13\u2^2-65\u2-335)/\nonumber\\
 &&(2*(\t2^3\u2-4
      \t2^3-3\t2^2\u2+12\t2^2+3\t2\u2-12\t2-\u2+4))
\end{eqnarray}
\begin{eqnarray}
      K_{22}&=&-(\s2^2\u2-4\s2^2+2\s2\t2-8\s2\u2+14\s2-2
      \t2+7\u2-106)/\nonumber\\
      &&(4*(\s2^4-4\s2^3+6\s2^2-4\s2+1))
\end{eqnarray}
\begin{eqnarray}
      K_{23}&=&-(\s2^2\t2-11\s2^2+\s2\t2^2+\s2\t2\u2-16\s2\t2-
      5\s2\u2+115\s2+\nonumber\\
      &&\t2^3-18\t2^2-\t2\u2^2-3\t2\u2+
      166\t2+13\u2^2-65\u2-335)/\nonumber\\
        &&(4(\s2^3\u2-4
      \s2^3-3\s2^2\u2+12\s2^2+3\s2\u2-12\s2-\u2+4))
\end{eqnarray}
\begin{eqnarray}
      K_{24}&=&(2\s2^2\t2+2\s2^2\u2-24\s2^2+\s2\t2^2+\s2\t2
      \u2-22\s2\t2-11\s2\u2+\nonumber\\
     &&  217\s2+\t2^2+7\t2\u2+16
      \t2-2\u2^2+41\u2-511)/(2*(\s2^2\t2\u2-4\s2^2\t2\nonumber\\
     && -\s2^2\u2+4\s2^2-2\s2\t2\u2+8\s2\t2+2\s2
      \u2-8\s2+\t2\u2-4\t2-\u2+4))
\end{eqnarray}
\begin{eqnarray}
      K_{33}&=&-(2\s2^2+2\s2\t2+6\s2\u2-46\s2+\t2^2\u2-10
      \t2^2-\nonumber\\
       &&4\t2\u2+74\t2+2\u2^3-22\u2^2+53\u2-118)/
 (\s2^2\u2^2-8\s2^2\u2+\nonumber\\
   && 16\s2^2-2\s2\u2
      ^2+16\s2\u2-32\s2+\u2^2-8\u2+16)
\end{eqnarray}
\begin{eqnarray}
      K_{34}&=&-(2*(\s2^2+\s2\t2-11\s2+\t2^2+2\t2\u2-19\t2+\u2
      ^3-9\u2^2+28\u2+11))/\nonumber\\
       &&(\s2\t2\u2^2-8\s2
      \t2\u2+16\s2\t2-\s2\u2^2+8\s2\u2-16\s2\nonumber\\
       &&-\t2\u2^2+8
      \t2\u2-16\t2+\u2^2-8\u2+16)
\end{eqnarray}
\begin{eqnarray}
      K_{44}&=&-(\s2^2\u2-10\s2^2+2\s2\t2-4\s2\u2+74\s2+2
      \t2^2+6\t2\u2-\nonumber\\
       &&46\t2+2\u2^3-22\u2^2+53\u2-118
      )/ (\t2^2\u2^2-8\t2^2\u2+16\t2^2-2\t2\u2\nonumber\\
&& ^2+16\t2\u2-32\t2+\u2^2-8\u2+16)
\end{eqnarray}

The differential cross section for subprocess $\gamma c\to J/\psi c$
can be written as follows:
\begin{equation}
 \frac{d\hat\sigma}{d\hat t}=\frac{1}{16\pi(\hat s-m^2)^2}|\bar M|^2
  \end{equation}
The total cross section will be obtained after integration over  $\hat
t$ in limits:
 $$\hat t_{max\atop{min}}=
    m^2-\frac{\hat s-m^2}{2\hat s}[\hat s-3m^2\pm
    \sqrt{(\hat s-9m^2)(\hat s-m^2)}].$$
 This procedure can be made analytically, we find that:
\begin{equation}
  \hat\sigma(\gamma c\to J/\psi c)=\frac{B_{\gamma c}}
   {16\pi(\hat s-m^2)^2}\sum_{j\geq i=1}^{4}[H_{ij}(\tilde s,\tilde t
_{max})-H_{ij}(\tilde s,\tilde t_{min})],
 \end{equation}
where $\tilde   t_{max}=\hat   t_{max}/m^2$   and   $\tilde   t_{min}=\hat
t_{min}/m^2$.
 The explicit expressions for functions  $H_{ij}(\tilde s,\tilde t)$
are more  unwieldy  than  for  functions  $K_{ij}$  and  the   FORTRAN
expression for $H_{ij}$ can be obtained by E-mail from the author on
request \footnote{saleev@univer.samara.su}.

We shall  also  calculate  the dominant  contribution   in   total   and
differential inclusive $J/\psi$ photoproduction cross sections from
$\gamma g\to J/\psi g$ partonic subprocess.
Here we present main formula for this one without discussion (see, for
example \cite{6}).

First the partonic differential cross section is given by:
\begin{equation}
 \frac{d\hat\sigma}{d\hat t}(\gamma g\to J/\psi g)=
  B_{\gamma g}M_J^4F(\hat s,\hat t),
\end{equation}
 where
 $$B_{\gamma g}=\frac{8\pi\alpha_s^2\Gamma_{ee}}{3\alpha M_J},$$
$$F(\hat s,\hat t)=\frac{1}{\hat s^2}
\left[ \frac{ \hat s^2(\hat s-M_J^2)^2+\hat t^2(\hat t-M_J^2)^2
  +\hat u^2(\hat u-M_J^2)^2}
{(\hat s-M_J^2)^2(\hat t-M_J^2)^2(\hat u-M_J^2)^2}\right].$$
Here: $\hat s+\hat t+\hat u=M_J^2$, $\hat t_{max}=0$ É $\hat t_{min}=
-\hat s+M_J^2.$

The total partonic cross section reads:
\begin{equation}
 \hat\sigma(\gamma g\to J/\psi g)=B_{\gamma g}M_J^4\Phi(\hat s),
\end{equation}
 where
 $$ \Phi(\hat s)=\frac{2}{(\hat s+M_J^2)^2}\left[\frac{\hat s-M_J^2}
     {\hat s M_J^2}-\frac{2\ln(\hat s/M_J^2)}{\hat s+M_J^2}
       \right]+$$
 $$\frac{2(\hat s+M_J^2)}{\hat s^2M_J^2(\hat s-M_J^2)}-
    \frac{4\ln(\hat s/M_J^2)}{\hat s(\hat s-M_J^2)^2}.$$

\section{Associated $J/\psi$  plus charm photoproduction on proton}
 Let us consider the kinematic for process $\gamma p\to J/\psi+c+X$ in
the rest frame of the proton (the lab. frame). Variables of partonic
subprocess $\gamma c\to J/\psi c$
and variables describing measurable process are connected as follows:
$$\hat s=xs+m^2\mbox{, }\hat t=5m^2-xzs\mbox{, }\hat u=-xs(1-z),$$
where (in lab. frame) $s=2m_pE_{\gamma}$, $z=E_J/E_{\gamma}$,
$x$ is a momentum fraction of charm quark in the proton.
  In the general factorization approach of QCD the measurable cross
section $\sigma$ and partonic cross section $\hat\sigma$ are connected
by the following expressions:
\begin{equation}
 \sigma(\gamma p\to J/\psi c X)=
  \int_{x_{min}}^{1}dxC_p(x,Q^2)\hat\sigma(\gamma c\to J/\psi c),
\end{equation}
 where $x_{min}=8m^2/s$, $Q^2=M_J^2$ and
\begin{equation}
 \frac{d\sigma}{dz}(\gamma p\to J/\psi cX)=
  -s\int dx xC_p(x,Q^2)\frac{d\hat\sigma}{d\hat t}(\gamma c\to J/\psi c),
   \end{equation}
 where the region of integration over $x$ is defined by the condition:
  $\hat t_{min}<5m^2-xzs<\hat t_{max}$.

At present the  direct  experimental  information  about  charm  quark
distribution function in a proton in wide region of $x$ is practically
absent. Existing parameterizations of
$C_p(x,Q^2)$ are very different. For comparison we shall use in our
calculation "hard" scaling parameterization \cite{1}:
 \begin{equation}
 C_p(x,Q^2)=C_p(x)=18x^2[(1-x)(1+10x+x^2)/3+2x(1+x)\ln x]
 \end{equation}
 and "soft",  based on perturbative QCD,  parameterization \cite{10} at
the scale $Q^2=M_J^2$:
\begin{equation}
 xC_p(x,Q^2)=(s-s_c)^a(1+Bx)(1-x)^D\exp\left (-E+\sqrt{E's^b\ln(1/x)}\right ),
\end{equation}
 where
 $$a=1.01,\quad b=0.37,\quad s_c=0.888,\quad B=4.24-0.804s,
   \quad D=3.46+1.076s,$$
 $$ E=4.61+1.49s,\quad E'=2.555+1.961s,\quad \mu^2=0.25\quad GeV^2,
  \quad \Lambda=0.232\quad GeV,$$
 $$ s=\ln\left(\frac{\ln(Q^2/\Lambda^2)}{\ln(\mu^2/\Lambda^2)}\right).$$

Note that mean value of the proton momentum,  which  is  carried by charm
quarks, is equal to 0.3\% in the case of parameterization \cite{1}
and approximately 0.5\% in the case of parameterization \cite{10},
which does not contradict data from EMC Collaboration on $F_2^c(x,Q^2)$
\cite{11}.

These parameterizations  for  $C_p(x,Q^2)$  have  very  different
physical interpretation.   The   scaling   parameterization   \cite{1},
so-called "intrinsic",  was obtained  assuming  existence of
 a small, but finite, non-perturbative charm component in proton
wave function.
It is  necessary  for  description  of open charm and $J/\psi$ production
total cross section and $x_F$ spectra at $x_F\to 1$ in hadron-hadron
and hadron-nucleus collisions. On the contrary, the parameterization
\cite{10}, so-called "extrinsic", strongly depends on choice of scale
$Q^2$, because  it  was obtained in perturbative QCD approach assuming
that at $Q^2<Q^2_{min}$ charm quarks in the proton are absent and they are
generated in QCD cascade only at large $Q^2$.

In Fig.2  the  $x$  dependence  of  the charm distribution function is
shown at $Q^2=M_J^2$. As it will be discussed later  different $x$-
dependences of the charm distribution functions give us very different
predictions for total $J/\psi$  photoproduction  cross  section  as  a
function of energy.

\section{Results and discussion}
 Result of calculation of  total  cross  section  as  function  of
photon energy  $E_{\gamma}$  for  associated  $J/\psi$ plus open charm
photoproduction in approach discussed above is shown in Fig.3.
"Intrinsic"  parameterization  \cite{1}  predicts largest
value of cross section at energies $E_{\gamma}\le 60$ GeV, but this cross
section rapidly decrease as energy growth at $E_{\gamma}\ge 100$ GeV.
On the contrary, the cross section, which was calculated using "extrinsic"
parameterization \cite{10},  monotonously  increase from $5\dot 10^{-3}$ nbn at
the
energy $E_{\gamma}=50$ GeV to 0.4 nbn at the energy  $E_{\gamma}=5$  TeV
and at  the  energy  range  of  $ep$-collider  HERA  ($\sqrt s_{\gamma
p}=200$ GeV) it is approximately equal to 0.8 nbn (Fig.4).  Such a  way
we have  sufficiently  large  measurable cross section of the $J/\psi$
plus open charm production for direct experimental  investigation of the
$x$-dependence of charm quark distribution function in a proton.

The study  of the $J/\psi$ photoproduction in $\gamma p$- interactions
concentrated on inelastic $J/\psi$ production  as  a  tool  to  obtain
information on the gluon distribution function in a proton \cite{5}.
That is why we have calculated the contribution of  the  $\gamma  c\to
J/\psi c$   partonic   subprocess  in  the  total  inclusive  $J/\psi$
photoproduction cross section.  Fig.4 shows result of calculation this
contribution using  "extrinsic"  parameterization  \cite{10} as well as
contribution of  the  dominant  photon-gluon  fusion   subprocess   as
functions of  $\sqrt  s_{\gamma  p}$.  It  is  known that $\gamma g\to
J/\psi g$ partonic subprocess gives contribution
in the total inclusive $J/\psi$ photoproduction cross section
  approximately  equal
to 50\%  at  the  energy  $\sqrt  s_{\gamma p}=200$ GeV \cite{5}.  Our
calculation shows that total contribution of the  subprocesses  $\gamma
c\to J/\psi c$ and $\gamma\bar c\to J/\psi\bar c$ at the energy $\sqrt
s_{\gamma p}=200$ GeV is  equal  to  4\%  of the photon-gluon fusion
subprocess contribution
or 2\% of the total contribution of all
mechanisms.  The contribution of the charm quarks in $J/\psi$
photoproduction cross section may be significant  and approximately equal to
contributions of elastic (5\%) or diffractive (2\%) mechanisms \cite{5}.
But at smaller energies $\sqrt s_{\gamma p}=10-20$ GeV charm
quark and untiquark contribution is 15\% of photon-gluon fusion
contribution in the case of parameterization \cite{1} for charm quarks
in the proton.

The contribution of the proton intrinsic charm in $J/\psi$ photoproduction
must be very large in the region of proton fragmentation that is at small
 $z=E_J/E_{\gamma}$ in the proton rest frame. Figs. 5 and 6 show
  results of calculation for contributions  $\gamma c\to J/\psi c$ and
  $\gamma g\to J/\psi g$ partonic subprocesses in $z-$spectra of the $J/\psi$
    photoproduction at energies $\sqrt s_{\gamma p}=14.7$ GeV and
 $\sqrt s_{\gamma p}=200$ GeV, correspondingly. The charm quark contribution
is larger than photon-gluon fusion subprocess contribution at $z<0.2$ for
both energies. At the relatively small energy
($\sqrt s_{\gamma p}=14.7$ GeV) the
parameterization \cite{1} gives more large contribution. At the energy
$\sqrt s_{\gamma p}=200$ GeV the contribution of parameterization \cite{1}
  is largest only for very small $z< 10^{-3}$, in the region
  $10^{-3}<z<0.2$ the contribution of parameterization \cite{10}
   is dominant.

In conclusion we note that the comparison of our results with data didn't made
because data usually have been obtained in fixed kinematical region of
variables $z$ and $p_T$\footnote{The introduction of the normalization
K-faktor is needed too}. In contrary in our approach total
 kinematical region (It is bounded by conservation laws)
of these variables have been took
into account. But all calculations were made in the same kinematical
approximation and we can to compare relative contributions of the different
mechanisms with data. The calculations for total cross section, $z$ and
$p_T$ spectra of the $J/\psi$ photoproduction via partonic subprocess
 $\gamma c\to J/\psi c$ in fixed kinematical region (HERA collider, for
   example) will be present in future publications.

I would like to thank Prof. N.Zotov and Dr. A.Martynenko for
useful discussions and particularly Prof. A.Likhoded for remarkes.
The work was supported in part by the Russia
Foundation for Basic Research under Grant 93-02-3545.
\end{section}


{\large\bf Figure captions}
\begin{enumerate}
\item Diagrams used to describe the partonic subprocess $\gamma c\to J/\psi c$.

\item The charm distribution function in a proton versus $x$ at $Q^2=M_J^2$.
   The curver 1 corresponds to parameterization \cite{1}, the curve 2 -
\cite{10}.

\item The cross section of the process $\gamma p\to J/\psi Ó X$ as a function
 of photon energy  $E_{\gamma}$. Curves as the same Fig.2.

\item The cross section of the process $\gamma p\to J/\psi X$ as a function of
  $\sqrt s_{\gamma p}$. The curver 1 corresponds to subprocess  $\gamma c\to
J/\psi c$
(parameterization for $C_p(x,Q^2)$ is taken from \cite{10}), the curver 2 -
$\gamma g\to J/\psi g$
 (parameterization for $G_p(x,Q^2)$ is taken from \cite{10} too).

\item The $z$- spectrum of $J/\psi$ in $\gamma p$- interaction
at $\sqrt s_{\gamma p}=14.7$ çÜ÷. Curvers 1 and 2 - contributions of
$\gamma c\to J/\psi c$ subprocess used parameterizations \cite{1} and
\cite{10},
  correspondingly.
The curver 3 - contribution of photon-gluon fusion mechanism.

\item As the same Fig. 5 at  $\sqrt s_{\gamma p}=200$ GeV.
\end{enumerate}

\newpage
\begin{center}
 {\large\bf Abstract}
\end{center}
Based on a perturbative theory of quantum chromodynamics and non-relativistic
quark model, associated $J/\psi$ plus open charm photoproduction on charm
quarks in a proton via partonic subprocess $\gamma c\to J/\psi c$ is discussed.
It is shown that the value and energy dependence of the cross section for
such process remarkably depends on the choice of charm distribution
function in a proton. In the region of small $z=E_J/E_{\gamma}<0.2$ the
contribution of the $\gamma c\to J/\psi c$ subprocess in the inelastic
$J/\psi$ photoproduction spectra is larger than the contribution  of the
 photon-gluon fusion subprocess. At the energy range of HERA collider
charm quarks contribution in the total inclusive $J/\psi$ photoproduction
cross section may be equal to 4\% of the dominant contribution of
photon-gluon fusion mechanism.
\newpage

\def\emline#1#2#3#4#5#6{%
       \put(#1,#2){\special{em:moveto}}%
       \put(#4,#5){\special{em:lineto}}}
\unitlength=1.00mm
\special{em:linewidth 1pt}
\linethickness{1pt}
\begin{picture}(125.00,94.20)
\emline{24.00}{54.20}{1}{34.00}{64.09}{2}
\emline{34.00}{64.09}{3}{34.00}{64.09}{4}
\emline{34.00}{64.09}{5}{49.00}{64.09}{6}
\put(50.00,69.04){\oval(2.00,9.89)[]}
\emline{50.00}{73.99}{7}{34.00}{73.99}{8}
\emline{34.00}{73.99}{9}{34.00}{83.88}{10}
\emline{34.00}{83.88}{11}{34.00}{83.88}{12}
\emline{34.00}{83.88}{13}{51.00}{83.88}{14}
\emline{34.00}{73.99}{15}{32.00}{71.84}{16}
\emline{32.00}{71.84}{17}{34.00}{71.84}{18}
\emline{34.00}{71.84}{19}{34.00}{71.84}{20}
\emline{34.00}{71.84}{21}{32.00}{70.12}{22}
\emline{32.00}{70.12}{23}{32.00}{70.12}{24}
\emline{32.00}{70.12}{25}{34.00}{70.12}{26}
\emline{34.00}{70.12}{27}{34.00}{70.12}{28}
\emline{34.00}{70.12}{29}{32.00}{67.96}{30}
\emline{32.00}{67.96}{31}{32.00}{67.96}{32}
\emline{32.00}{67.96}{33}{34.00}{67.96}{34}
\emline{34.00}{67.96}{35}{34.00}{67.96}{36}
\emline{34.00}{67.96}{37}{32.00}{66.24}{38}
\emline{32.00}{66.24}{39}{32.00}{66.24}{40}
\emline{32.00}{66.24}{41}{34.00}{66.24}{42}
\emline{34.00}{66.24}{43}{34.00}{66.24}{44}
\emline{34.00}{66.24}{45}{32.00}{64.09}{46}
\emline{32.00}{64.09}{47}{32.00}{64.09}{48}
\emline{32.00}{64.09}{49}{34.00}{64.09}{50}
\emline{51.00}{70.12}{51}{57.00}{70.12}{52}
\emline{57.00}{70.12}{53}{51.00}{70.12}{54}
\emline{51.00}{70.12}{55}{57.00}{70.12}{56}
\emline{51.00}{67.96}{57}{57.00}{67.96}{58}
\emline{56.00}{66.24}{59}{60.00}{68.82}{60}
\emline{60.00}{68.82}{61}{60.00}{68.82}{62}
\emline{60.00}{68.82}{63}{55.00}{71.84}{64}
\emline{24.00}{94.20}{65}{26.00}{92.05}{66}
\emline{28.00}{89.90}{67}{30.00}{88.18}{68}
\emline{32.00}{86.03}{69}{34.00}{83.88}{70}
\put(32.00,91.19){\makebox(0,0)[cc]{k}}
\put(32.00,54.20){\makebox(0,0)[cc]{q}}
\put(56.00,86.89){\makebox(0,0)[cc]{$q'$}}
\put(44.00,77.86){\makebox(0,0)[cc]{p}}
\put(44.00,58.93){\makebox(0,0)[cc]{p}}
\put(72.00,68.82){\makebox(0,0)[cc]{$p_J=2p$}}
\emline{92.00}{73.98}{71}{102.00}{73.98}{72}
\emline{92.00}{73.98}{73}{82.00}{64.09}{74}
\emline{82.00}{83.87}{75}{84.00}{82.15}{76}
\emline{84.00}{82.15}{77}{84.00}{82.15}{78}
\emline{86.00}{80.86}{79}{88.00}{79.14}{80}
\emline{88.00}{79.14}{81}{88.00}{79.14}{82}
\emline{90.00}{76.99}{83}{92.00}{74.84}{84}
\emline{102.00}{73.98}{85}{104.00}{71.83}{86}
\emline{104.00}{71.83}{87}{104.00}{73.98}{88}
\emline{104.00}{73.98}{89}{106.00}{71.83}{90}
\emline{106.00}{71.83}{91}{106.00}{73.98}{92}
\emline{106.00}{73.98}{93}{108.00}{71.83}{94}
\emline{108.00}{71.83}{95}{108.00}{73.98}{96}
\emline{108.00}{73.98}{97}{110.00}{71.83}{98}
\emline{110.00}{71.83}{99}{110.00}{73.98}{100}
\emline{110.00}{73.98}{101}{112.00}{71.83}{102}
\emline{112.00}{71.83}{103}{112.00}{73.98}{104}
\emline{102.00}{73.98}{105}{109.00}{82.15}{106}
\emline{112.00}{73.98}{107}{119.00}{82.15}{108}
\put(114.00,81.94){\oval(10.00,2.15)[]}
\emline{112.00}{73.98}{109}{121.00}{67.10}{110}
\emline{113.00}{83.01}{111}{118.00}{89.04}{112}
\emline{116.00}{83.01}{113}{120.00}{88.18}{114}
\emline{115.00}{88.18}{115}{121.00}{91.19}{116}
\emline{121.00}{86.88}{117}{121.00}{91.19}{118}
\emline{22.00}{17.63}{119}{50.00}{17.63}{120}
\emline{33.00}{17.63}{121}{31.00}{19.78}{122}
\emline{31.00}{19.78}{123}{33.00}{19.78}{124}
\emline{33.00}{19.78}{125}{31.00}{21.93}{126}
\emline{31.00}{21.93}{127}{33.00}{21.93}{128}
\emline{33.00}{21.93}{129}{31.00}{23.65}{130}
\emline{31.00}{23.65}{131}{33.00}{23.65}{132}
\emline{33.00}{23.65}{133}{31.00}{25.80}{134}
\emline{31.00}{25.80}{135}{33.00}{25.80}{136}
\emline{33.00}{25.80}{137}{31.00}{27.52}{138}
\emline{31.00}{27.52}{139}{50.00}{27.52}{140}
\put(50.00,22.58){\oval(2.00,9.89)[]}
\emline{51.00}{23.65}{141}{57.00}{23.65}{142}
\emline{51.00}{20.64}{143}{57.00}{20.64}{144}
\emline{55.00}{25.80}{145}{60.00}{21.93}{146}
\emline{55.00}{18.92}{147}{60.00}{22.79}{148}
\emline{41.00}{27.52}{149}{39.00}{29.68}{150}
\emline{38.00}{30.54}{151}{36.00}{32.69}{152}
\emline{35.00}{33.55}{153}{33.00}{35.70}{154}
\emline{32.00}{36.56}{155}{30.00}{38.71}{156}
\emline{31.00}{27.52}{157}{50.00}{38.71}{158}
\emline{93.00}{22.80}{159}{91.00}{24.95}{160}
\emline{91.00}{24.95}{161}{91.00}{24.95}{162}
\emline{91.00}{24.95}{163}{93.00}{24.95}{164}
\emline{93.00}{24.95}{165}{93.00}{24.95}{166}
\emline{93.00}{24.95}{167}{91.00}{27.10}{168}
\emline{91.00}{27.10}{169}{91.00}{27.10}{170}
\emline{91.00}{27.10}{171}{93.00}{27.10}{172}
\emline{93.00}{27.10}{173}{93.00}{27.10}{174}
\emline{93.00}{27.10}{175}{91.00}{28.82}{176}
\emline{91.00}{28.82}{177}{93.00}{28.82}{178}
\emline{93.00}{28.82}{179}{91.00}{30.97}{180}
\emline{91.00}{30.97}{181}{93.00}{30.97}{182}
\emline{93.00}{30.97}{183}{91.00}{32.69}{184}
\emline{91.00}{32.69}{185}{91.00}{32.69}{186}
\put(112.00,27.75){\oval(2.00,9.89)[]}
\emline{113.00}{28.82}{187}{120.00}{28.82}{188}
\emline{113.00}{25.81}{189}{120.00}{25.81}{190}
\emline{120.00}{25.81}{191}{120.00}{25.81}{192}
\emline{118.00}{24.09}{193}{125.00}{27.10}{194}
\emline{118.00}{30.97}{195}{125.00}{27.10}{196}
\emline{93.00}{22.80}{197}{112.00}{22.80}{198}
\emline{112.00}{32.69}{199}{82.00}{32.69}{200}
\emline{102.00}{32.69}{201}{100.00}{34.84}{202}
\emline{99.00}{35.70}{203}{97.00}{37.85}{204}
\emline{96.00}{38.71}{205}{94.00}{40.86}{206}
\emline{93.00}{41.72}{207}{91.00}{43.87}{208}
\emline{93.00}{22.80}{209}{106.00}{13.77}{210}
\end{picture}
\begin{center}Fig.~1
\end{center}
\end{document}